\documentclass[MSNbibl,nameyear,dvips]{arxstspdf}
\usepackage{flushend}
\usepackage{stfloats}
\volume{29}
\issue{2}
\pubyear{2014}
\firstpage{261}
\lastpage{266}
\doi{10.1214/14-STS482} % kopijuoti is 'New paper accepted'
\referstodoi{10.1214/13-STS457}% pagrindinio straipsnio DOI, kai straipsnis yra diskusija ar rejoinder'is



\begin{document}
\begin{frontmatter}

\vspace*{6pt}\title{Rejoinder: ``On the Birnbaum Argument for the Strong Likelihood Principle''}%\thanksref{T1}
% kai straipsnis turi susijusiu diskusiju ir rejoinder'iu
\runtitle{Rejoinder}

\begin{aug}
\author[a]{\fnms{Deborah G.}~\snm{Mayo}\corref{}\ead[label=e1]{mayod@vt.edu}}
\runauthor{D. G. Mayo}

\affiliation{Virginia Tech}

\address[a]{Deborah G. Mayo is Professor of Philosophy, Department of Philosophy,
Virginia Tech, 235 Major Williams Hall, Blacksburg, Virginia 24061, USA
\printead{e1}.}
\end{aug}

% ABSTRACT

% KEYWORDS
% Pirmas kwd is didziosios raides
\end{frontmatter}

%s1 #&#
\section{Introduction}\label{sec1}

I am honored and grateful to have so many interesting and challenging
comments on my paper. I want to thank the discussants for their willingness
to jump back into the thorny quagmire of Birnbaum's argument. To a question
raised in the paper ``Does it matter?'', these discussions show the answer
is yes. The enlightening connections to contemporary projects are
especially valuable in galvanizing future efforts to address foundational
issues in statistics.

As long-standing as Birnbaum's result has been, Birnbaum himself went
through dramatic shifts in a short period of time following his famous
(1962) result. More than of historical interest, these shifts provide a
unique perspective on the current problem. Already in the rejoinder to
\citet{Bir62}, he is worried about criticisms (by \cite*{Prat62}) pertaining
to applying WCP to his constructed mathematical mixtures (what I call
Birnbaumization), and hints at replacing WCP with another principle
(Irrelevant Censoring). Then there is a gap until around 1968 at which
point Birnbaum declares the SLP plausible ``only in the simplest case,
where the parameter space has but two'' predesignated points [\citet{Bir68}, page
301]. He tells us in Birnbaum (\citeyear{Bir70N1}, page 1033) that he has pursued the
matter thoroughly, leading to ``rejection of both the likelihood concept
and various proposed formalizations of prior information.'' The basis for
this shift is that the SLP permits interpretations that ``can be seriously
misleading with high probability'' [\citet{Bir68}, page 301]. He puts forward the
``confidence concept'' (Conf) which takes from the Neyman--Pearson (N--P)
approach ``techniques for systematically appraising and bounding the
probabilities (under respective hypotheses) of seriously misleading
interpretations of data'' while supplying it an evidential interpretation
[\citet{Bir70N1}, page 1033]. Given the many different associations with
``confidence,'' I~use (Conf) in this Rejoinder to refer to Birnbaum's idea.
Many of the ingenious examples of the incompatibilities of SLP and (Conf)
are traceable back to Birnbaum, optional stopping being just one [see
\citet{Bir69}]. A bibliography of Birnbaum's work is \citet{Gie77}. Before
his untimely death (at 53), Birnbaum denies the SLP even counts as a
principle of evidence (in \cite*{Bir77}). He thought it anomalous that
(Conf) lacked an explicit evidential interpretation, even though, at an
intuitive level, he saw it as the ``one rock in a shifting scene'' in
statistical thinking and practice [\citet{Bir70N1}, page 1033]. I return to
this in Section~\ref{sec4} of this rejoinder.

%s2 #&#
\section{Bj{\o}rnstad, Dawid and Evans}\label{sec2}

Let me begin by answering the central criticisms that, if correct, would be
obstacles to what I purport to have shown in my paper. It is entirely
understandable that leading voices in a long-lived controversy would assume
that all of the twists and turns, avenues and roadways, have already been
visited, and that no new flaw in the argument could enter to shake up the
debate. I say to the reader that the surest sign that the issue is
unsettled is that my critics disagree among themselves about the puzzle and
even the key principles under discussion: the WCP, and in one case, the SLP
itself. To remind us [Section~2.2]:
\begin{quote}
SLP: For any two experiments $E_{1}$ and $E_{2}$
with different probability models $f_{1}$, $f_{2}$ but
with the same unknown parameter $\theta$, if outcomes $\mathbf{x}^{\ast}$ and
$\mathbf{y}^{\ast}$ (from $E_{1}$ and $E_{2}$, resp.)
determine the same likelihood function [$f_{1}(\mathbf{x}^{\ast}; \theta) =
cf_{2}(\mathbf{y}^{\ast}; \theta)$ for all $\theta$], then $\mathbf{x}^{\ast}$ and
$\mathbf{y}^{\ast}$ should be inferentially equivalent for any inference
concerning parameter $\theta$.
\end{quote}
A shorthand for the entire antecedent is that ($E_{1}, \mathbf{x}^{\ast}$) is
an \textit{SLP pair} with ($E_{2}, \mathbf{y}^{\ast}$), or just $\mathbf{x}^{\ast}$
and $\mathbf{y}^{\ast}$ form an SLP pair (from
$\{E_{1},\mathrm{E}_{2}\}$). Assuming all the SLP stipulations, we have
\begin{quote}
SLP: If ($E_{1}, \mathbf{x}^{\ast}$) and ($E_{2}, \mathbf{y}^{\ast}$) form an SLP
pair, then $\operatorname{Infr}_{E_1}[ \mathbf{x} ^{\ast}]= \operatorname{Infr}_{E_2}[\mathbf{y} ^{\ast}]$.
\end{quote}

\subsection*{Bj{\o}rnstad}

According to Bj{\o}rnstad, ``The starting point is \textit{not} that we have
an arbitrary outcome of one single experiment, but rather that two
experiments have been performed about the same parameter resulting in
proportional likelihoods.'' I do not think Bj{\o}rnstad can actually mean to
say the SLP cannot be applied until both members of the SLP pair are
observed. So, for example, if in the sequential experiment one is able to
stop (with a 0.05 $p$-value) at $n = 169$, resulting in $\mathbf{y}^{\ast}$, one may
not regard it as evidentially equivalent to $\mathbf{x}^{\ast}$, the SLP pair
with $n$ fixed at 169, until and unless $\mathbf{x}^{\ast}$ is actually
generated? The universal generalization of the SLP asserts that for an
arbitrary $\mathbf{y}^{\ast}$, if it has an SLP pair $\mathbf{x}^{\ast}$, then
$\mathbf{y}^{\ast}$ is equivalent in evidence to $\mathbf{x}^{\ast}$. Bj{\o}rnstad's
problematic reading results in his next remark: ``So Birnbaum does not
enlarge a known single experiment but constructs a mixture of the two
performed experiments.'' What is constructed in Birnbaum's experiment
$E_{B}$ is a \textit{hypothetical or mathematical
mixture}, based on having observed $\mathbf{y}^{\ast}$ (from $E_{2}$).
This is part of the key gambit I call \textit{Birnbaumization} (Section~2.4). We are to consider the possibility that performing $E_{2}$
(which gave rise to $\mathbf{y}^{\ast}$) was the result of a $\theta$-irrelevant
randomizer (deciding between $E_{1}$ or $E_{2}$). Now I
grant Birnbaum that we may imagine all the SLP pairs are ``out there,''
each pair assumed to have resulted from a $\theta$-irrelevant randomizer,
ripe for plucking whenever a member of an SLP pair is observed. (See
Sections~2.5 and 5.1.) Yet even granting Birnbaum all of this, we still may
not infer SLP (nor does it follow in the case where the mixture is
actual).

Bj{\o}rnstad also criticizes me because he claims the SLP ``is simply
\textit{not about method evaluation}.'' His position is that there is an
evidential appraisal, and quite separately an assessment of long-run
performance. For a frequentist, or one who holds Birnbaum's (Conf),
evidential import is inseparable from an assessment of the relevant error
probabilities. Not because we regard evidential import as all about
long-runs, but because scrutinizing a given inference is bound up with a
method's ability to have alerted us to misleading interpretations.

Bj{\o}rnstad does ``not find any \textit{new} clarification of Birnbaum's
fundamental theorem in this paper'' because he assumes I am channeling the
attempts of \citet{Dur70}, \citet{Kal75N2} and \citet{EvaFraMon86N3}, all of whom restrict or modify either SP or WCP to block the
result. While I stand on the shoulders of these and other earlier
treatments, a crucial difference is that, unlike them, I do not alter the
principles involved. If one is out to demonstrate the logical flaw in an
argument, as every good philosopher knows, one should scrupulously adhere
to the premises and generously interpret the machinations of the arguer.
This I do. Bj{\o}rnstad's opinion is that ``one may regard the paper by Mayo
as actually not discussing the LP at all.'' Or, alternatively, one may
regard the position held by this critic to be mistaken about the SLP and
Birnbaum's argument.

\subsection*{Dawid}

Professors Dawid and Evans disagree about the key principle invoked in
Birnbaum's argument, the WCP. Dawid views it as an equivalence relation,
Evans says it is not. I follow Birnbaum in regarding the WCP as an
equivalence, but, unlike both Dawid and Evans, I pin down what is to be
meant in regarding WCP as an equivalence, or, for that matter, an
inequivalence (see Section~4.3). First Dawid.

Dawid maintains that my WCP differs from the principle of conditionality
Birnbaum uses in the SLP argument. Not so. I am working with the WCP stated
in Birnbaum (\citeyear{Bir62}, \citeyear{Bir69}), the very same one defined by Dawid:
\begin{quote}
The evidential meaning of any outcome of any mixture experiment is the same
as that of the corresponding outcome of the corresponding component
experiment, ignoring the over-all structure of the mixture experiment.
\end{quote}
Dawid's definition is a portion of the one found in \citet{Bir62}, page
271. It assumes, of course, all of the other stipulations, for example, we
are making ``informative'' inferences about $\theta$, it is a
$\theta$-irrelevant mixture, the outcome is given, and all the rest. It is
the definition used in countless variations of the SLP argument, and it is
clearly captured in my Section~4.3. Perhaps I should have abbreviated it as CP; WCP
comes from Berger and Wolpert's (\citeyear{BerWol84}) manifesto, \textit{The}
\textit{Likelihood Principle}. My intention was to underscore Birnbaum's
emphasis that the WCP concerns mixture experiments and is distinct from
many other uses of ``conditioning'' in statistics [\citet{Bir62}, pages
282--283].

I wondered why Dawid thought I denied that WCP asserts an equivalence,
until I noticed that Dawid lops off the end of my sentence from Section~7:
I do not say ``the problem stems from mistaking WCP as the equivalence''
simpliciter, but rather it stems from the incorrect equivalence! The
incorrect equivalence equates the inference from the given experiment with
one that takes account of the (irrelevant) mixture structure. This is what
Dawid is on about in describing invalid construals of WCP, so he can
scarcely object. (See Section~5.2.)

As with any equivalence, there is an implicit inequivalence as a corollary.
[See (i) and (ii) in Section~4.3.] Typically, in saying the evidential import of two
outcomes are the same, one would not add ``and be sure to ignore any
features that would render them inequivalent.'' Birnbaum adds this warning
because some treatments do not ignore the mixture structure. To put this
another way, WCP includes the phrases ``is the same as'' as well as
``ignoring.'' The problem is that Dawid is ignoring the word ``ignoring''
in the very definition he proffers. There is no difference between the
phrases
\begin{itemize}
\item[$\bullet$] ignore the over-all structure of the mixture experiment
\end{itemize}
and
\begin{itemize}
\item[$\bullet$] eschew any construal that does not ignore the over-all
structure of the mixture experiment.
\end{itemize}

\hspace*{-1pt}I also refer to this as irrelevance (Irrel) (Section~4.3.2) because Birnbaum
describes the WCP as asserting the ``irrelevance of (component) experiments
not actually performed'' [\citet{Bir62}, page 271].

Dawid opines that I am using the WCP in David Cox's (\citeyear{Cox58}) famous weighing
example, which he does not define; I am guessing he means to suggest I must
be limiting myself to actual mixtures. That is to miss the genius of
Birnbaum's argument. Birnbaum, quite deliberately, intends to capitalize on
the persuasiveness of conditioning in Cox's famous example, but his ploy is
to extend the argument to mathematical or hypothetical\vadjust{\goodbreak} mixtures. (I am not
saying it is an innocuous move, but that is a separate matter.) Even if
Dawid chooses to view Cox's WCP as a nonequivalence, it is irrelevant; I am
following Birnbaum in construing it as an equivalence, permitting, for
example, $\mathbf{y}^{\ast}$, known to have come from a nonmixture, to be
evidentially equivalent to the appropriate $\theta$-irrelevant mixture as
in Section~4.3. (Irrel) protects against illicit readings that Dawid warns against.
SLP still will not follow.

So Dawid, Birnbaum and I are using the same definition of WCP. The onus is
on Dawid to pinpoint where my characterization deviates from Birnbaum's.
The only difference is that I have shown one cannot get to SLP, and Dawid
gives no clue how to get around my criticism.
%Curiously, Dawid maintains
%that if we restrict WCP to actual mixtures, then he agrees with me that SLP
%does not follow. Yet if the argument for the SLP is retained, why would it
%fail for the case of an actual SLP pair, as in the example of Section~5.4?

For Dawid to simply pronounce that ``Birnbaum's theorem is indeed logically
sound'' and that therefore my argument ``must itself be unsound'' is
question-begging, and will not do. Demonstrating unsoundness of my argument
should be accomplished straightforwardly, as I have done regarding
Birnbaum. That said, I fully agree with Dawid that one can view [(SP and
WCP), entails SLP] as a theorem, but in order to \textit{detach} the SLP,
as is mandatory for Birnbaum, he is left with a ``proof'' that is either
unsound or question-begging. Perhaps those who are long wedded to
Birnbaum's argument are comfortable with merely assuming what was to have
been shown. It is part of the mysterious ``path of enlightenment followed
by conversion'' that Dawid mentions. That is no reason for others to allow
``trust me, it is sound'' to take the place of argument.

\subsection*{Evans}

Given that Evans largely agrees with me, it may seem ungenerous to focus on
apparent disagreements, but there is too great a danger in leaving some
misimpressions regarding a problem already beset with decades of
misunderstanding. Notably, it seems I have not convinced Evans of the
logical error that Birnbaum makes. Instead Evans thinks the problem is with
the conditionality principle WCP, and claims that frequentists need to fix
it somehow. But it is not the principle, it is the ``proof.''

I have at least convinced Evans that there are cases where SP and WCP and
not-SLP hold without logical contradiction (in \cite*{autokey26}). These cases may
be called ``counterexamples'' to the argument whose conclusion is the SLP.
They are also counterexamples to [WCP entails SLP], using the weaker
notion\vadjust{\goodbreak}
of mathematical equivalence of \citet{Bir72} that dispenses with (SP).
Evans will take those counterexamples to show that WCP is not an
equivalence relation, assuming a frequentist standpoint. Now it is true
that any such counterexample may be seen to warn us against mistaking WCP
as asserting the incorrect equivalence, noted in my rejoinder to Dawid. But
that does not preclude WCP from asserting a correct equivalence. A~more
general issue I have with Evans' treatment is that it does not show where
the source of the problem lies in arguments for SLP. Introducing his
set-theoretic treatment into a simple argument, I am afraid, does not help
to pinpoint where the argument goes wrong, but in fact leaves us with a
very murky idea even as to his definition of WCP. The argument for the SLP
begins with: We are given $\mathbf{y}^{\ast}$ from $E_{2}$, a member of
an SLP pair. Will Evans block introduction of the mathematical mixture in
Birnbaumization? This would seem to cut off Birnbaum's argument too
quickly. Were that sufficient, the debate would have surely ended with
\citet{Kal75N2}. Note too, unlike Evans, my argument in the paper under
discussion does not rely on assuming a frequentist principle at all, though
obviously I avoid a formulation that rules it out in advance. To sum up
this section, Evans uses my counterexamples to show a restricted WCP may be
applied, while blocking SLP. Left as it is, it opens him to the criticism
(the one Dawid raised!) that he is altering Birnbaum's WCP and restricting
it to actual mixtures.

What a surprise, then, to hear Evans allege that ``many authors, including
Mayo, refer to the [WCP] which restricts attention to ancillaries that are
physically part of the sampling.'' I do not know on what grounds Evans
wants to distinguish actual and mathematical mixtures, but Birnbaum's
argument for the SLP concerns mathematical or hypothetical mixtures.
Birnbaum calls an experiment a mixture ``if it is mathematically
equivalent'' to a mixture [\citet{Bir62}, page 279]. Further, \citet{Bir62}
emphasizes that earlier proofs [that WCP and SP imply SLP] were restricted
to actual mixtures. ``But in the above proof'' he is able to get a result
relevant for all classes of experiments by using an ancillary ``constructed
with the hypothetical mixture'' [$E_{B}$] [\citet{Bir62}, page 286]. So, I
am not sure what Evans is alleging. In one place, Evans worries whether the
WCP ``resolves the problem with conditionality more generally,'' but this
is a separate issue from Birnbaum's argument. Here the focus is on WCP
solely for purposes of arriving at the SLP.\vadjust{\goodbreak}

Although there was not space to discuss this in my paper, it is worth
noting why merely blocking the SLP with a modified WCP fails to make
progress with a further goal required of an adequate treatment. Consider
how, in discussing Durbin's modified principle of conditionality, Birnbaum
notes that ``Durbin's formulation (C'), although weaker than [WCP], is
nevertheless too strong (implies too much of the content of [SLP]) to be
compatible with standard (non-Bayesian) statistical concepts and
techniques'' [\citet{Bir70N2}, page 402]. Birnbaum (\citeyear{Kal75N1}, page 264) raises the same
problem with Kalbfleish's restriction to ``minimal experiments'' to which
Evans' treatment is closely related. Evans does not show his modified
conditionality principle avoids entailing ``too much of the SLP.'' (This
relates to Dawid's point about stopping rules in his comment.) For a
frequentist account to satisfy Birbaum's (Conf), all cases that allow
misleading interpretations with high probability should still show up as
SLP violations.

To this end, my argument shows that any violation of SLP in frequentist
sampling theory necessarily results in an illicit substitution in the
formulation of Birnbaum's argument. To put the problem in general terms,
$p = r$ does not follow from $p = q$
and $q = r$, if $q$ shifts to $q'$ within the argument, where
$q \neq q'$ (fallacy of 4 terms). For specifics see Section~5.
Thus, ours is in no danger of implying ``too much'' of the SLP: what was an
SLP violation remains one. Now Evans may not be concerned with retaining
those frequentist SLP violations, given he makes it very clear he embraces
Bayesianism, but that is irrelevant to what an adequate treatment of
Birnbaum's argument demands. I have seen some statistics textbooks leave
the details of the SLP proof to the reader; I think it is time to give full
credit to students who found it impossible to make a valid substitution in
general. I explained why.

%s3 #&#
\section{Fraser, Hannig, Martin and Liu}\label{sec3}

Let me turn to the second group of discussants. It is an honor to be
``strongly commended'' by Fraser for emphasizing the importance of
``principles and arguments of statistical inference''; and I feel my
efforts are worthwhile with Martin and Liu's noting my ``demonstration
resolves the controversy around Birnbaum and LP, helping to put the
statisticians' house in order.'' I entirely agree with them that the
``confusion surrounding Birnbaum's claim has perhaps
discouraged\vadjust{\goodbreak}
researchers from considering questions about the foundations of
statistics,'' at least from appealing to those foundations that reject the
SLP. Let me underscore Fraser's point that the need for an inferential
variation of (N--P) theory ``reached the mathematical statistics community
rather forcefully with \citet{Cox58}; this had the focus on the two
measuring-instruments example and on uses of conditioning that were compelling.''
Cox's (\citeyear{Cox58}) example also appears in Hannig's discussion, and I will borrow
his simple description of the case where the measurement but not the
instrument $M$ is observed. In that case, inference is based on the convex
combination of the mixture components, consistent with WCP. This allows me
to succinctly put an equivocation that I suspect may enter, in the case of
SLP pairs, between the irrelevance of the mixture structure, given
($E_{i}, \mathbf{z}_{i}$), and the irrelevance of the
index $i$, given just the measurement. This equivocation may be behind the
Birnbaum puzzle.

Fraser rightly reminds us that, ``statistical inference as an alternative
to (N--P) decision theory has a long history in statistical thinking'' with strong
impetus from Fisher. Still Birnbaum struggled to articulate a N--P theory as
``an inference theory'' (\cite{Bir77}), and my view is that we had to
solve ``Birnbaum's problem'' before doing so properly. Finding Birnbaum's
argument unsound opens the door to foundations that are free from paying
obeisance to the SLP. In this spirit Martin and Liu correctly view my paper
as ``an invitation for a fresh discussion on the foundations of statistical
inference.'' Yet there is more than one way of proceeding. Tracing out the
mathematical similarities and differences between the approaches of Fraser,
Hanning, Martin and Liu is a task for which others are better equipped than
I. All are said to violate SLP.

It is interesting to note, as Hannig does, that ``since the mid 2000s,
there has been a true resurrection of interest in modern modifications of
fiducial inference'' which had long fallen into disrepute. Fraser's has
been one of the leading voices to persevere with innovative developments,
and his own ``confidence'' idea is clearly in sync with Birnbaum. However,
the differences that emerge in this group's discussions should not be
downplayed. Hannig says that ``the common thread for these approaches is a
definition of inferentially meaningful probability statements about subsets
of the parameter space without the need for subjective prior information,''
and Martin and Liu suggest that error probability accounts are appropriate
only for decision procedures, as distinct from their ``inferential
models.'' Some might view these as attempts to build a concept of evidence
as a kind of \textit{probabilism} but without the priors. However,
in the background of these contemporary developments lurks a suspicion that
their SLP violations were picking up differences where no purely
inferential difference was warranted. So long as Birnbaum's proof stood,
this suspicion made sense.

Post-SLP, it is worth standing back and reflecting anew on these accounts.
In this respect, this foundational project is just beginning because for 40
or 50 years, the questions of foundations were largely restricted to
accounts that obeyed, or were close to obeying, the SLP. So, we have
Birnbaum, alongside Fisher, being catapulted onto the contemporary
foundational scene, squarely calling on us to address the still unresolved
problem: how to obtain an account of statistical inference that also
controls the probability of seriously misleading inferences. Better yet,
the two goals should mesh into one.

%s4 #&#
\section{Post-SLP Foundations}\label{sec4}

Return to where we left off in the opening section of this rejoinder:
\citet{Bir69},
\begin{quote}
The problem-area of main concern here may be described as that of
determining precise \textit{concepts of statistical evidence}
(systematically linked with mathematical models of experiments), concepts
which are to be \textit{non-Bayesian, non-decision-theoretic}, and
significantly \textit{relevant to statistical practice}. [\citet{Bir69},
page 113.]
\end{quote}
Given Neyman's behavioral decision construal, Birnbaum claims that ``when a
confidence region estimate is interpreted as representing statistical evidence about a
parameter'' [\citet{Bir69}, page 122], an investigator has necessarily adjoined a
concept of evidence, (Conf) that goes beyond the formal theory. What is
this evidential concept? The furthest Birnbaum gets in defining (Conf) is
in his posthumous article \citet{Bir77}:
\begin{quote}
(Conf) A concept of statistical evidence is not plausible unless it finds
`strong evidence for $H_{2}$ against $H_{1}$' with small probability
($\alpha$) when $H_{1}$ is true, and with much larger probability ($1 -
\beta$) when $H_{2}$ is true. [\citet{Bir77}, page 24.]
\end{quote}
On the basis of (Conf), Birnbaum reinterprets statistical outputs from N--P
theory as strong, weak, or worthless statistical evidence depending on the
error probabilities of the test [\citet{Bir77}, pages 24--26]. While this sketchy
idea requires extensions in many ways (e.g., beyond pre-data error
probabilities and beyond the two hypothesis setting), the spirit of (Conf),
that error probabilities quantify properties of methods which in turn
indicate the warrant to accord a given inference, is, I think, a valuable
shift of perspective. This is not the place to elaborate, except to note
that my own twist on Birnbaum's general idea is to appraise evidential
warrant by considering the capabilities of tests to have detected erroneous
interpretations, a concept I call \textit{severity}. That Birnbaum
preferred a propensity interpretation of error probabilities is not
essential. What matters is their role in picking up how features of
experimental design and modeling alter a methods' capabilities to control
``seriously misleading interpretations.'' Even those who embrace a version
of probabilism may find a distinct role for a severity concept. Recall that
Fisher always criticized the presupposition that a single use of
mathematical probability must be competent for qualifying inference in all
logical situations [\citet{Fis56}, page 47].

Birnbaum's philosophy evolved from seeking concepts of evidence in degree
of support, belief or plausibility between statements of data and
hypotheses to embracing (Conf) with the required control of misleading
interpretations of data. The former view reflected the logical empiricist
assumption that there exist context-free evidential relationships---a
paradigm philosophers of statistics have been slow to throw off. The newer
(post-positivist) movements in philosophy and history of science were just
appearing in the 1970s. Birnbaum was ahead of his time in calling for a
philosophy of science relevant to statistical practice; it is now long
overdue!
\begin{quote}
``Relevant clarifications of the nature and roles of statistical evidence
in scientific research may well be achieved by bringing to bear in
systematic concert the scholarly methods of statisticians, philosophers and
historians of science, and substantive scientists...'' [\citet{Bir72}, page
861].
\end{quote}

% zodis "Acknowledgments" paliekamas pagal autoriu

%suskaldyti doi

% imsref loaded by jurgita.kaciuliene, 2014-05-29 09:16:16
% imsref loaded by jurgita.kaciuliene, 2014-05-29 09:41:53
% imsref loaded by jurgita.kaciuliene, 2014-05-29 09:43:34
% imsref loaded by jurgita.kaciuliene, 2014-05-29 09:47:11
% imsref loaded by jurgita.kaciuliene, 2014-05-29 09:59:27
% imsref loaded by jurgita.kaciuliene, 2014-05-29 10:00:07

\end{document}